\documentclass[aps,pra,twocolumn,amsmath,superscriptaddress,floatfix]{revtex4}

\usepackage{amsmath}
\usepackage{epsfig}
\usepackage{graphicx}
\usepackage{graphics}
\usepackage{url}
\usepackage[english]{babel}
\usepackage{dcolumn}
\usepackage{braket}
\usepackage{color}

\newcommand{\cre}[1]{\hat{#1}^{\dagger}}
\renewcommand{\bf}[1]{\mathbf{#1}}
\begin{document}
\title{Spatiotemporal dynamics of Coulomb-correlated carriers in semiconductors}

\author{F.~Lengers}
\affiliation{Institut f\"ur Festk\"orpertheorie, Universit\"at M\"unster,
Wilhelm-Klemm-Str.~10, 48149 M\"unster, Germany}

\author{R.~Rosati}
\affiliation{Chalmers University of Technology, Department of Physics, 412 96 Gothenburg, Sweden}

\author{T.~Kuhn}
\affiliation{Institut f\"ur Festk\"orpertheorie, Universit\"at M\"unster,
Wilhelm-Klemm-Str.~10, 48149 M\"unster, Germany}

\author{D.~E.~Reiter}
\affiliation{Institut f\"ur Festk\"orpertheorie, Universit\"at M\"unster,
Wilhelm-Klemm-Str.~10, 48149 M\"unster, Germany}

\date{\today}

\begin{abstract}
When the excitation of carriers in real space is focused down to the nanometer scale, the carrier system can no longer be viewed as homogeneous and ultrafast transport of the excited carrier wave packets occurs. In state-of-the-art semiconductor structures like low-dimensional heterostructures or monolayers of transition metal dichalcogenides, the Coulomb interaction between excited carriers becomes stronger due to confinement or reduced screening. This demands a fundamental understanding of strongly interacting electrons and holes and the influence of Coulomb correlations. To study the corresponding particle dynamics in a controlled way we consider a system of up to two electron-hole pairs exactly within a wave function approach. We show that the excited wave packets contain a non-trivial mixture of free particle and excitonic states. We further scrutinize the influence of Coulomb interaction on the wave packet dynamics revealing its different role for below and above band-gap excitation.
\end{abstract}

\maketitle

\section{Introduction}

\begin{figure}[h!]
\includegraphics[width=0.7\columnwidth]{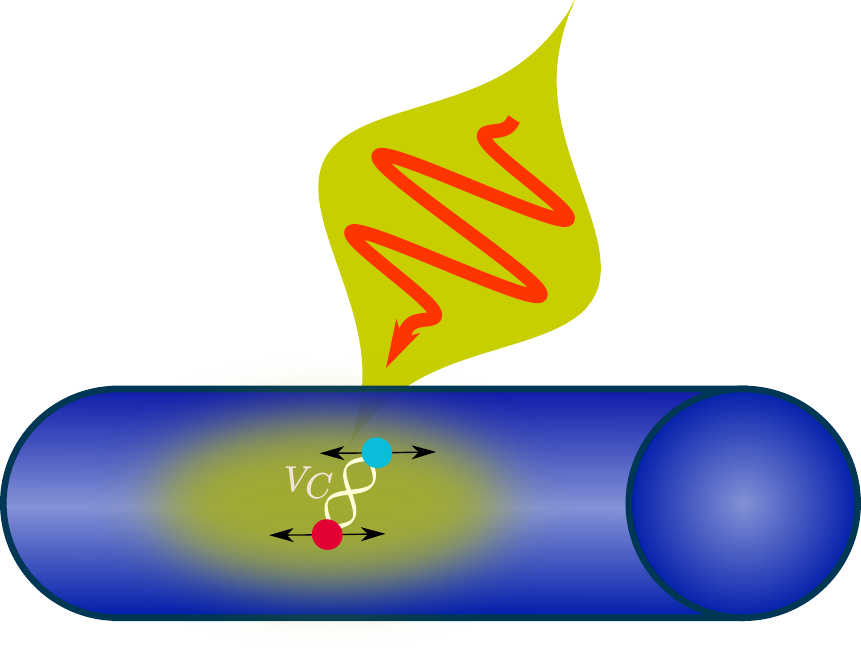}
\caption{Sketch of localized photoexcitation of a quantum wire resulting in traveling electron-hole pairs coupled via the Coulomb interaction $V_C$.}
\label{fig_1}
\end{figure}

Research of exciton-based devices is a growing field due to the promising aspect of direct interconnection of electronic signal processing and optical communication \cite{Lienau00,Vasa09,High08,Grosso09,Violante14}. At the heart of most of these devices is the ultrafast motion of photoexcited electrons and holes within the semiconductor accessible by various experimental techniques \cite{Guenther02,Vasa09,Man16}. The Coulomb interaction between electrons and holes in such devices is becoming more decisive, because in state-of-the-art semiconductor structures like low-dimensional heterostructures or monolayers of transition metal dichalcogenides, the Coulomb interaction is enhanced due to the confinement and excitonic effects are becoming more pronounced. Exciton physics has additionally been boosted by the discovery of strongly bound excitons in monolayers of transition metal dichalcogenides \cite{Mak10,Splendiani10,Wang18,Mueller18} and recently the spatiotemporal dynamics of these strongly bound excitons are being explored \cite{Kulig18,Cadiz18,Jin18}. Therefore, it is of crucial importance to understand the impact of the Coulomb interaction on the ultrafast localized excitation and dynamics of interacting carrier wave packets. Such localized excitation can then be used to investigate fundamental processes like charge transfer and capture processes which naturally occur on nanometric scales. In this paper, we give a detailed view on this aspect in an exact description for the correlated carrier dynamics.\\
%% \\
%
%
In view of nanometric length scales and ultrashort time scales, we are entering scales where semiclassical descriptions are not able to properly describe physical processes \cite{Rossi02}. While the ultrafast carrier dynamics including coherent and incoherent excitons has been extensively studied for homogeneous excitations \cite{Vu00,Siantidis01,Kira06,Brem18}, inhomogeneous treatments are more complicated. The aspect of spatiotemporal dynamics of photoexcited carriers on ultrashort time- and length scales has been studied either in the limit of low densities where Hartree-Fock approximations are applicable \cite{Steininger96,Herbst03,Pasenow05,Reiter06,Reiter07, Lengers17} or in the limit of an exact number of carriers taken as an initial condition \cite{Grasselli15,Grasselli16}. While the treatment of the low-density case revealed the dynamics of the carrier excitation and the fundamental carrier and polarization transport after excitation, the latter underlines the effects of carrier correlations leading to strong deviations from the free-carrier behavior when treating correlated particles. Here, we use a theory based on a wave function approach. By restricting ourselves to excitations of up to two electron-hole pairs, we are able to treat the carrier system and their Coulomb correlations exactly. We further account for the interaction of carriers with a strongly localized light field. In our model, we observe the transition between free-carrier transport and correlated carrier transport. To study a computationally feasible problem, which at the same time exhibits strong Coulomb effects, we use an optically excited one-dimensional (1D) semiconductor quantum wire as sketched in Fig.~\ref{fig_1}.\\
The paper is organized as follows: In Sec.~\ref{sec_theory} we define the Hamiltonian describing the carrier system including Coulomb- and light field-interaction. With this we set up our wavefunction ansatz as well as the equations of motion. In Sec.~\ref{sec_results_low_dens} we consider a low-intensity excitation both below and above the band gap. In Sec.~\ref{sec_results_high_dens} we then increase the excitation power and accordingly the mean number of carriers, resulting in a different dynamics. To evaluate the influence of the excitons, in Sec.~\ref{sec_excitons} we analyze the results in the excitonic picture interpolating between  low and high excitation. We finish with conclusions in Sec.~\ref{sec_concl}.
\section{Theory}\label{sec_theory}
\subsection{Hamiltonian} \label{sec_theory_ham}
As a system we assume a CdTe quantum wire with a $100\,\mathrm{nm^2}$ cross-section as sketched in Fig.\ref{fig_1}. We restrict ourselves to the lowest carrier subbands, i.e. one conduction and one valence band, which we assume to be spin-degenerate. The Hamiltonian of the non-interacting system is described by
	\begin{align*}
	\hat{H}_0=\sum\limits_{k,\sigma}\epsilon_{k}^e \cre{c}_{k\sigma}\hat{c}_{k\sigma}+\sum\limits_{k,\sigma}\epsilon_{k}^h \cre{d}_{k\sigma}\hat{d}_{k\sigma}
	\end{align*}
with $\hat{c}_{k\sigma}(\cre{c}_{k\sigma})$ being the electron and $\hat{d}_{k\sigma}(\cre{d}_{k\sigma})$ the hole annihilation (creation) operators of a state with spin $\sigma$ ($=\pm 1/2$) and longitudinal wavevector $k$; the energies are $\epsilon_{k}^e=E_{\mathrm{gap}}+\frac{\hbar^2}{2m_e}k^2$ and $\epsilon_{k}^h=\frac{\hbar^2}{2m_h}k^2$ with the effective masses $m_e=0.091m_0$ and $m_h=0.41m_0$ ($m_0$ being the free electron mass) \cite{La82}.\\
The system is excited by a linearly polarized localized laser pulse. The carrier-light-field interaction in dipole approximation reads
	\begin{align*}
	\hat{H}_{\mathrm{cf}}=-\sum\limits_{\substack{k,k' \\ \sigma}}\left[E_{k',k}\cre{c}_{k'\sigma}\cre{d}_{-k\bar{\sigma}}+c.c.\right],
	\end{align*}
where $E_{k',k}(t)=\tilde{\mathbf{E}}(k-k',t)\cdot \mathbf{M}_{k',k}$ is the spatial Fourier transform of the electric field amplitude $\bf{E}(z,t)$ multiplied by the interband dipole matrix element $\mathbf{M}_{k',k}$, which is assumed to be independent of $k$. Due to linear polarization the dipole matrix element can be taken as independent of spin and spin is conserved with $\bar{\sigma}=-\sigma$.\\
The Coulomb interaction in our system is \cite{Rossi02}
	\begin{align*}
		\hat{H}_C=\frac{1}{2}\sum\limits_{\substack{k,k',q \\ \sigma,\sigma'}} &V_q\left[
	\cre{c}_{k\sigma}\cre{c}_{k'\sigma'}\hat{c}_{k'+q\sigma'}\hat{c}_{k-q\sigma}\right. \\
	& +\cre{d}_{k\sigma}\cre{d}_{k'\sigma'}\hat{d}_{k'+q\sigma'}\hat{d}_{k-q\sigma}\\
&\left.-2\cre{c}_{k\sigma}\cre{d}_{k'\sigma'}\hat{d}_{k'+q\sigma'}\hat{c}_{k-q\sigma}\right]\,.
	\end{align*}
 The first (second) term describes the electron-electron (hole-hole) interaction, while the third term is the attractive electron-hole interaction, which eventually will give rise to excitonic effects. For the Coulomb matrix element $V_q$ we use the bulk matrix element with static dielectric constant $\epsilon_s=10.5$ multiplied by the quantum wire form factor \cite{Herbst03}.\\
\subsection{Equations of motion}
The excitation of the quantum wire should be such that only up to two electron hole-pairs can be excited. This allows us to scrutinize the effects of the Coulomb interaction using a well defined
wave function ansatz
	\begin{align}
	\ket{\Psi}=
	& a^{(0)}\ket{0}+\sum\limits_{\substack{k,k' \\ \sigma,\sigma'}}a_{\substack{k,k'\\ \sigma,\sigma'}}^{(1)}\cre{c}_{k\sigma}\cre{d}_{k'\sigma'}
	\ket{0}
	\label{eq_wavefunction_ansatz} 
	\\
	& +\sum\limits_{\substack{k_1,k_2,k_3,k_4 \\ \sigma_1,\sigma_2,\sigma_3,\sigma_4}}
	a_{\substack{k_1,k_2,k_3,k_4 \\ \sigma_1,\sigma_2,\sigma_3,\sigma_4}}^{(2)}	
	\cre{c}_{k_1\sigma_1}\cre{c}_{k_2\sigma_2}
	\cre{d}_{k_3\sigma_3}\cre{d}_{k_4\sigma_4}\ket{0}\nonumber \,.
	\end{align}
This wave function describes a state composed of one and two electron hole pairs and the electron-hole vacuum $\ket{0}$ via the wave function coefficients $a^{(0)}, a^{(1)}, a^{(2)}$. The validity of our approach can be tuned by the laser power and pulse duration since these parameters control the amount of excited carriers. Note that consistent with an initially undoped semiconductor, we assume electron-hole symmetry. This ansatz corresponds to a Configuration-Interaction (CI) approach often used in quantum chemistry for ground-state calculations of $N$-electron systems, although with our excitation-controlled electron-hole density we circumvent the usual problems of a truncated CI treatment \cite{Axt98,Bartlett07} . Remarks on the comparison of the dynamical CI treatment and density-matrix approaches are given in App.~\ref{app_theory}.\\ %While the CI approach is exact for an $N$-electron system when considering up to $N$-particle excitations (full CI), our approach treats a system with $2$ excited electron-hole pairs exactly. 
Next, we set up the equations of motion for the wave function coefficients $a^{(0)}, a^{(1)}, a^{(2)}$ using the Schr\"odinger equation
	\begin{align*}
	i\hbar\frac{d}{dt}\ket{\Psi}=\hat{H}\ket{\Psi}.
	\end{align*}
Inserting the wave function ansatz, commuting the fermionic annihilation operators to the right and multiplying from the left with a 0-,1- or 2-pair state  ($\bra{0}$, $\bra{0}\hat{d}_1\hat{c}_2$ or $\bra{0}\hat{d}_1\hat{d}_2\hat{c}_3\hat{c}_4$) leads then to the equations of motion, where the indices refer to the combination of wave vector and spin. To illustrate the occuring expectation values, we evaluate the most complicated expectation value as an example: 
	\begin{align}
	&\bra{0}\hat{d}_1\hat{d}_2\hat{c}_3\hat{c}_4\cre{c}_{\alpha}\cre{c}_{\beta}
	\cre{d}_{\gamma}\cre{d}_{\delta}\ket{0}=
	\label{eq_expectationvalue}\\
	&\delta_{\alpha,4}\delta_{\beta,3}(\delta_{\gamma,2}\delta_{\delta,1}-\delta_{\gamma,1}\delta_{\delta,2})
	-\delta_{\alpha,3}\delta_{\beta,4}(\delta_{\gamma,2}\delta_{\delta,1}-\delta_{\gamma,1}\delta_{\delta,2}).
	\nonumber
	\end{align}
The $\delta$-expressions give the same coefficients of the wave function because of the fermionic-induced symmetries 
	\begin{align*}
	a^{(2)}_{1234}=-a^{(2)}_{2134}=a^{(2)}_{2143}=-a^{(2)}_{1243}.
	\end{align*}
Due to the spin conservation only coefficients with equal number of up and down spins can occur. We therefore define
	\begin{align}
	a^{(2)}_{\sigma,\sigma,\bar{\sigma},\bar{\sigma}}
	&=:a^{(2),P}\\
	a^{(2)}_{\sigma,\bar{\sigma},\bar{\sigma},\sigma}
	&=:a^{(2),A}
	\end{align}
being the coefficients for parallel ($a^{(2),P}$ describing electrons (holes) with the same spin) and antiparallel ($a^{(2),A}$ describing electrons (holes) with different spin) carrier states. Further identifying 
	\begin{align}
	a^{(1)}_{\sigma_1,\sigma_2}=:a^{(1)}\delta_{\sigma_1,\bar{\sigma_1}}
	\end{align}
 we may drop all spin-indices in the following.\\
The equation of motion for the 0-pair coefficient reads
	 \begin{align}
 i\hbar\frac{d}{dt}a^{(0)}=-2\sum\limits_{\substack{k,k'}}E_{-k',k}^* a_{\substack{k,k'}} ^{(1)}, \notag
 \end{align}
showing that only the coupling to a 1-pair coefficient due to an electric field results in a change of $a^{(0)}$. \\
The equation for the 1-pair coefficients is
 \begin{align}
 i\hbar\frac{d}{dt}a_{\substack{k,k'}}^{(1)}=&
 	\left(\epsilon_{k}^e+\epsilon_{k'}^h\right)a_{\substack{k,k' }}^{(1)}
 	-\sum\limits_q V_q a_{\substack{k-q,k'+q }}^{(1)}  \notag \\
 	& -E_{-k',k}a^{(0)} \notag  \\
 	+4\sum\limits_{\substack{k_1,k_2 }}& \left[E_{-k_2,k_1}^* \left(a_{\substack{k,k_1,k',k_2 }}^{(2),P }+a_{\substack{k,k_1,k',k_2 }}^{(2),A}\right)\right]\,.\notag 
 	\label{eq_onepair_source}
 \end{align}
The first two terms on the right hand side describe the excitonic dynamics, noting that for $k'=-k$ they resemble the Wannier equation for direct excitons. The third and the last two terms are source terms resulting from the coupling to zero pairs and to two pairs, respectively. The prefactor $4$ of the last sum stems from the fermionic commutation relations.\\ % in (\ref{eq_expectationvalue}).\\
The equation of motion for the 2-pair coefficients $a^{(2),i}$ ($i=A,P$) are
 \begin{align}
 i\hbar\frac{d}{dt}a_{\substack{k_1,k_2,k_3,k_4 }}^{(2),i}=&
 \left[\epsilon_{k_1}^e+\epsilon_{k_2}^e+\epsilon_{k_3}^h+\epsilon_{k_4}^h
 \right]a_{\substack{k_1,k_2,k_3,k_4 }}^{(2),i} \notag \\
 &\hspace{-1cm}+\frac{1}{4}E_{-k_4,k_2}a_{\substack{k_1,k_3}}^{(1)}
 -\delta_{i,P}\frac{1}{4}E_{-k_3,k_2}a_{\substack{k_1,k_4}}^{(1)} \notag \\
 &\hspace{-1cm}-\delta_{i,P}\frac{1}{4}E_{-k_4,k_1}a_{\substack{k_2,k_3}}^{(1)}
 +\frac{1}{4}E_{-k_3,k_1}a_{\substack{k_2,k_4}}^{(1)} \notag \\
 &\hspace{-2cm}+\sum\limits_q
  V_q\left[a_{\substack{k_1-q,k_2+q,k_3,k_4}}^{(2),i} +
 a_{\substack{k_1,k_2,k_3-q,k_4+q }}^{(2),i}\right. \notag \\
&\hspace{-0.9cm}\left. -a_{\substack{k_1-q,k_2,k_3+q,k_4 }}^{(2),i}
 -a_{\substack{k_1,k_2-q,k_3+q,k_4 }}^{(2),i}\right. \notag \\
&\hspace{-0.9cm}\left. -a_{\substack{k_1-q,k_2,k_3,k_4+q }}^{(2),i}
 -a_{\substack{k_1,k_2-q,k_3,k_4+q }}^{(2),i}\right]\,. \notag 
 \end{align}
In this equation, the dynamics of the two-pair exciton (or biexciton) is given by the first term and last six terms, while the rest describe the excitation from the one-pair exciton.
We note that the equations of motion for $a^{(0)}$ and $a^{(1)}$ are exact, while in principle in the equation of motion for $a^{(2)}$ we would have a source term stemming from $a^{(3)}$. Because we have restricted ourselves to two eletron-hole pairs at most, this source term vanishes in our considerations.\\
The equations of motion are then solved by numerical integration with the initial condition of $a^{(0)}=1$ and  $a^{(1)}=a^{(2)}=0$ discretized on a $k$-space grid with 60 points.

\subsection{Dynamical quantities}
The spatiotemporal dynamics of the excited carriers is encoded in the space dependent density $n_{e/h}(z)$ for electrons and holes given by 
\begin{align}
	 n_{e}(z) =\Braket{\hat{n}_e(z)}=\frac{1}{V}\sum\limits_{{k},{k}',\sigma}\Braket{\cre{c}_{{k\sigma}}\hat{c}_{{k'\sigma}}}e^{i({k}'-{k}){z}} \quad
\end{align}
and the analogous definition for holes.
Here $z$ is the longitudinal position and the factor $2$ stems from the spin degeneracy due to linear polarization of the exciting electric field.
The expectation value for electrons is given by
 \begin{align}
 \Braket{\cre{c}_{k\sigma}\hat{c}_{k'\sigma}}=
 &\sum\limits_{k_1} a_{\substack{k',k_1 }}^{(1)}\cdot
	a_{\substack{k,k_1 }}^{(1)*}\nonumber\\
	+\sum\limits_{k_1,k_2,k_3}
	&4\left[
	2a_{k_1,k',k_2,k_3}^{(2),P}\cdot a_{k_1,k,k_2,k_3 }^{(2),P*}\right.\nonumber\\
	&\hspace{0.3cm}\left. 
	+4a_{\substack{k',k_1,k_3,k_2 }}^{(2),A}\cdot a_{k,k_1,k_3,k_2 }^{(2),A*} \notag
	\right],
 \end{align}
being independent of spin, and analog for the holes.\\
In order to further analyze the dynamical behavior of the energies in the system we define the  kinetic energies for electrons and holes
	\begin{align*}
	E_{\mathrm{kin}}^e
	&=\sum\limits_{{k,\sigma}}\left(\epsilon_{{k}}^e-E_{\mathrm{gap}}\right)\Braket{\cre{c}_{{k\sigma}}\hat{c}_{{k\sigma}}}\\
	E_{\mathrm{kin}}^h
	&=\sum\limits_{{k,\sigma}} \epsilon_{{k}}^h\Braket{\cre{d}_{{k\sigma}}\hat{d}_{{k}}}
	\end{align*}
and the interaction energy
	\begin{align*}
	E_{\mathrm{int}}=
	\frac{1}{2}\sum\limits_{\substack{k,k',q \\ \sigma,\sigma'}} V_q
&\left[\Braket{\cre{c}_{k\sigma}\cre{c}_{k'\sigma'}\hat{c}_{k'+q\sigma'}\hat{c}_{k-q\sigma}}\right.\\
 & \hspace{-3cm}\left.+\Braket{\cre{d}_{k\sigma}\cre{d}_{k'\sigma'}\hat{d}_{k'+q\sigma'}\hat{d}_{k-q\sigma}}
 -2\Braket{\cre{c}_{k\sigma}\cre{d}_{k'\sigma'}\hat{d}_{k'+q\sigma'}\hat{c}_{k-q\sigma}}\right]. 
	\end{align*}
The interaction energy can be further divided into the Hartree-Fock energy and the correlation energy. The Hartree-Fock energy is calculated by factorizing the two-particle into one-particle density matrices \cite{Rossi02}
	\begin{align*}
	E_{\mathrm{HF}}=
	\sum\limits_{{k,k',q}} 
	& V_q\left[
	2f_{{k'},{k'+q}}^ef_{{k},{k-q}}^e-f_{{k},{k'+q}}^ef_{{k',k-q}}^e\right.\\
	&\left.+2f_{{k',k'+q}}^h f_{{k,k-q}}^h-f_{{k,k'+q}}^h f_{{k',k-q}}^h\right.\\
	&\left.-4f_{{k',k'+q}}^ef_{{k,k-q}}^h
	-2p_{{k'+q,k}}^*p_{{k',k-q}}\right],
	\end{align*}
where $f^i$ are the electron (hole) intraband coherences for $i=e$ ($i=h$) and $p$ are the interband-polarizations (see App.~\ref{app_theory}).
With this we define the correlation energy as
	\begin{align}
	E_{\mathrm{cor}}=E_{\mathrm{int}}-E_{\mathrm{HF}} \,.
	\end{align}
This quantity gives us a direct measure for the correlation of the system. The correlations thereby describe the effects beyond simple electrostatic and exchange Coulomb-interactions of carriers. 

We remark that all energies shown in the following are normalized to the final mean number of electrons 
$N_e=2\sum\limits_k \Braket{\cre{c}_k\hat{c}_k}$.

\section{Results}\label{sec_results}

\begin{figure}[t]
\includegraphics[width=\columnwidth]{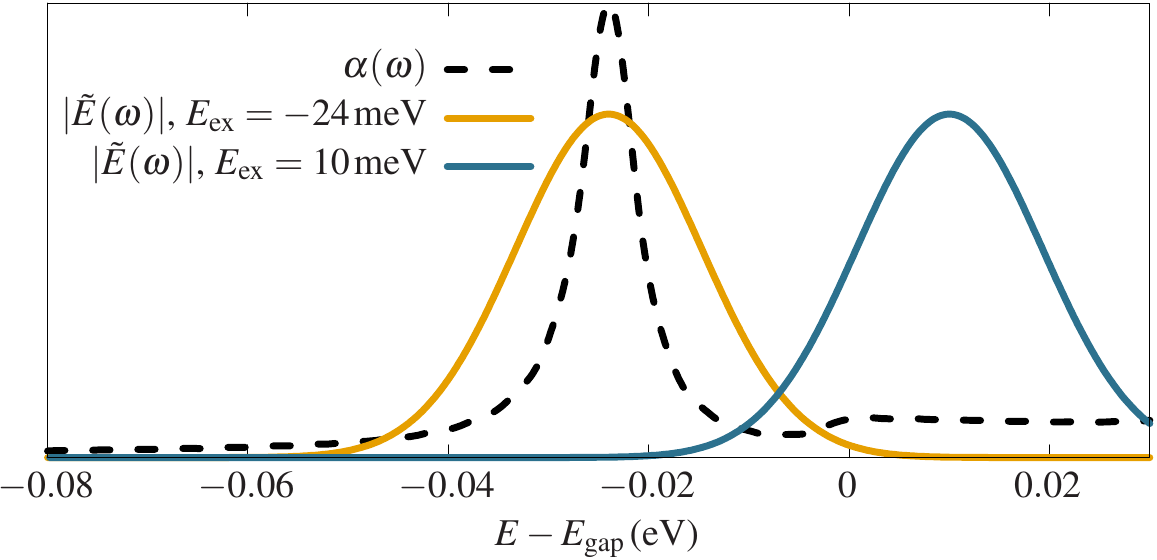}
\caption{Linear absorption spectrum (dashed line) and spectrum (solid lines) of the exciting laser pulses with different excess energy $E_{\mathrm{ex}}$.}
\label{fig_2}
\end{figure}
We now consider the dynamics of the optically generated carriers within the quantum wire. For the excitation we consider a pulse which is Gaussian in both space and time. We set the spatial variance of the pulse to $10\,$nm, corresponding to a full width at half maximum $\mathrm{FWHM}\approx 23.5\,$nm. The pulse duration is set to $100\,$fs. Further parameters are the excitation energy $\hbar \omega_L$ and the intensity, which we will vary in the following.\\
Our system can be characterized by its absorption spectrum $\alpha(\omega)$ as shown in Fig.~\ref{fig_2}, where a phenomenological dephasing time of $250\,$fs was added. The absorption spectrum can be divided into two parts, namely the excitonic resonance at $-24\,$meV below the band gap $E_{\mathrm{gap}}$ and the continuum states for energies above the band gap $E> E_{\mathrm{gap}}$. In the following we will consider two distinct excitation energies: (i) An excitation at the exciton resonance below the band gap, i.e.,  laser excess energy $E_{\mathrm{ex}}=\hbar \omega_L-E_{\mathrm{gap}}=-24$~meV, and (ii) an excitation into the continuum states with a laser excess energy $E_{\mathrm{ex}}=10$~meV. The corresponding laser pulse spectra are marked in Fig.~\ref{fig_2} with the excitonic excitation at $E_{\mathrm{ex}}=-24\,$meV as orange line and the continuum excitation with $E_{\mathrm{ex}}=10\,$meV as blue line. 

These two excitation conditions for low excitation strength will results in the excitation of either exclusively excitons or free carriers, respectively. For increasing field strength, we increase the particle number and we expect that Coulomb correlations will become important. Therefore, the excitation strength will be our tuning knob.
\subsection{Low-density limit}\label{sec_results_low_dens}
\begin{figure}[t]
\includegraphics[width=\columnwidth]{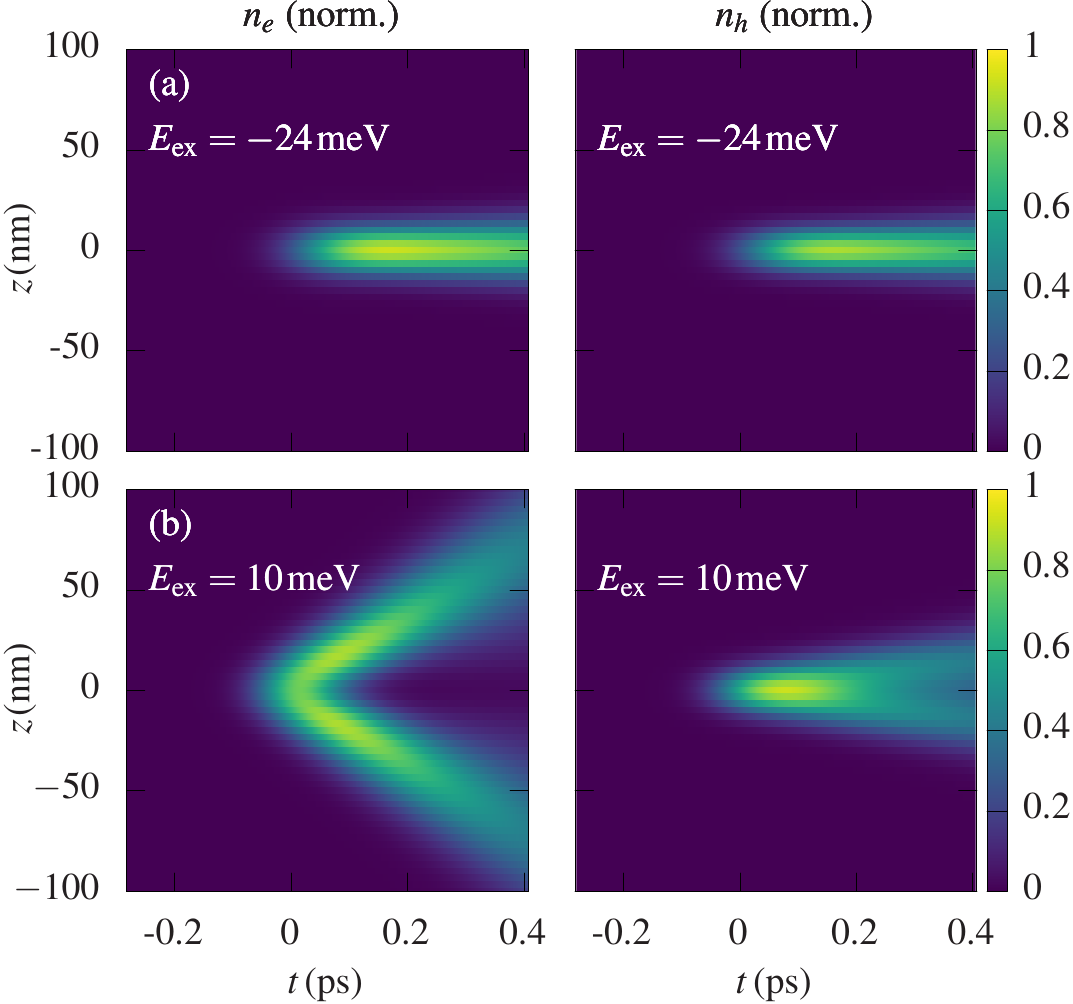}
\caption{Dynamics of the electron (left) and hole (right) density $n_{e/h}$ in the low-density limit for an excitation (a) resonant to the 1s exciton and (b) within the continuum. All the distributions have been normalized to their respective maxima.}
\label{fig_3}
\end{figure}

We start by looking at an excitation with a low laser power, such that the total number of electrons after the pulse is $N_e\approx 10^{-4}$, which we refer to as low-density limit.
In Fig.~\ref{fig_3} we plot the dynamics of electron (left) and hole (right) densities for an excitation (a) below and (b) above the band gap.\\
When exciting the system at the exciton resonance (Fig.~\ref{fig_3}(a)), we find an electron-hole pair at $z=0$ with almost no spatial dynamics. Here electrons and holes are bound by the Coulomb interaction in the 1s-exciton without any center-of-mass momentum. In contrast, when exciting within the continuum (Fig.~\ref{fig_3}(b)), independent electron and hole wave packets are excited which travel along the wire. %the electron and can see the carrier dynamics after excitation within the continuum of uncorrelated electron-hole pairs.
% This can be seen by the fact that independent electron and hole wave packets are excited which travel along the wire \cite{Steininger96}. 
Their velocity is dictated by the excess energy according to $v_e=\frac{\hbar k_0}{m_e}$ ($v_h=\frac{\hbar k_0}{m_h}$) for electrons (holes), where $k_0= \sqrt{2\mu E_{\mathrm{ex}}}/\hbar$, $\mu^{-1}=m_e^{-1}+m_h^{-1}$ being the reduced mass. Therefore the holes travel much slower and due to the low density the carrier wave packets do not affect each other. Our results in the low-density limit agree well with calculations for excitation in semiconductor quantum wells \cite{Steininger96}.

\begin{figure}[ht]
\includegraphics[width=\columnwidth]{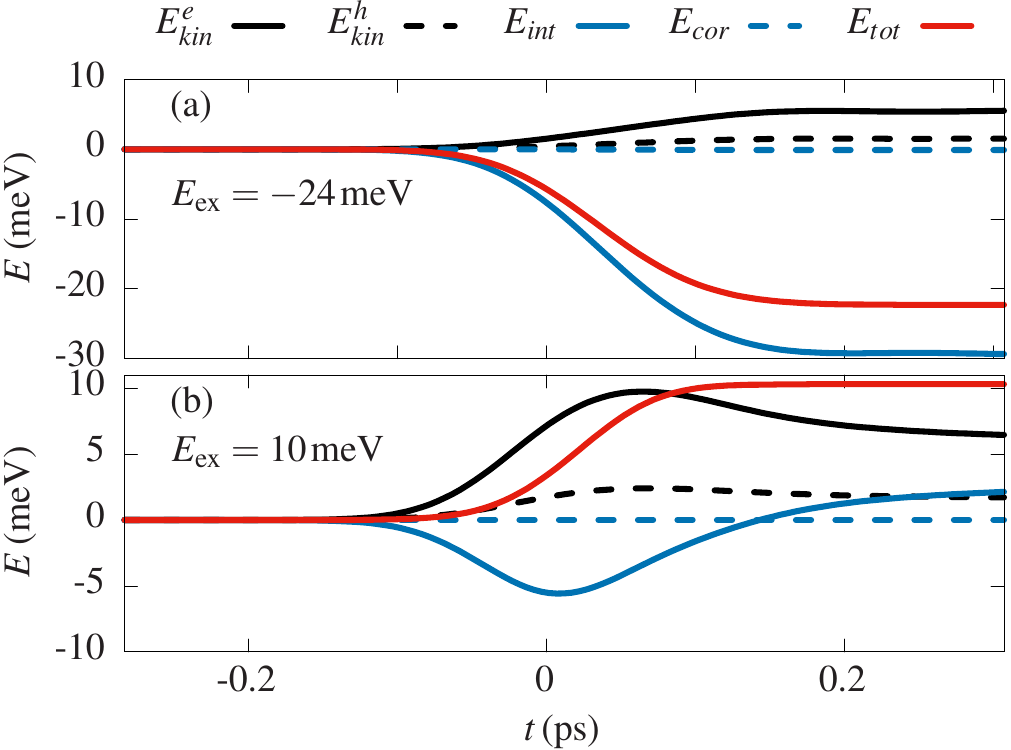}
\caption{Energy contributions normalized to the final electron number for an excitation (a) resonant to the exciton and (b) within the continuum.}
\label{fig_4}
\end{figure}

In Fig.~\ref{fig_4} we study the energy contributions normalized to the final electron number for the two excitation conditions, to get more insight into the Coulomb effects. For the excitation of the exciton, given in Fig.~\ref{fig_4}(a), we find that $E_{\mathrm{tot}}$ (red line) is close to the binding energy of the 1s exciton, underlining the fact that a bound electron-hole pair is excited. The kinetic energies for electrons and holes (black lines) have small values compared to the total energy. The strong influence of the Coulomb interaction is also seen in the interaction energy $E_{\mathrm{int}}$ (blue line), which is much stronger than the kinetic ones. The interaction energy mostly describes coherent excitons which are well-described in a Hartree-Fock picture and, accordingly, the correlation energy $E_{\mathrm{cor}}$ vanishes. This leads  to the conclusion that a Hartree-Fock picture is applicable, in which only coherent excitons are present, in agreement with the low-density limit. In this limit, the dynamics is still linear in the electric field $E$ and all many-particle quantities factorize like
	\begin{align*}
	\Braket{\cre{c}_1\cre{d}_2\hat{d}_3\hat{c}_4}=\Braket{\cre{c}_1\cre{d}_2}\Braket{\hat{d}_3\hat{c}_4}+\Braket{\cre{c}_1\hat{c}_4}\Braket{\cre{d}_2\hat{d}_3}+\mathcal{O}(E^4).
	\end{align*}
Here the first term on the right hand side is of order $E^2$ and already the second term is of order $E^4$ \cite{Axt94}.\\
The energies in the case of continuum excitation, displayed in Fig.~\ref{fig_4}(b), are completely different. Note the different scales of Fig.~\ref{fig_4}(a) and (b). The total energy approaches the excess energy of the exciting laser pulse, while the kinetic energies are similar to the below band-gap excitation. The main difference lies in the interaction energy, which starts off with negative values, showing that, even though we excited the system within the continuum, the Coulomb interaction alters the carrier dynamics directly above the band gap, in particular during the laser pulse. After the laser pulse the interaction energy approaches a small positive value one order of magnitude smaller than the interaction energy in Fig.\ref{fig_4}(a). If one increases the excess energy even further, a gradual decrease of the interaction energy occurs (not shown). 
 Again, due to the low-density limit, the correlation energy vanishes. 
\subsection{High-density limit}\label{sec_results_high_dens}
We will now turn to higher excited densities ($1\le N_e\le 2$) in which the electron and hole wave packets should interact strongly with each other. We will refer to this as high-density limit. While in the low-density case one could describe the $E_{\mathrm{ex}}=-24\,$meV and the $E_{\mathrm{ex}}=10\,$meV case by excitonic and free-carrier excitation, respectively, in a high-density case we expect a more complex picture, because the former interpretations rely on low-density eigenstates of the semiconductor.\\
\subsubsection{Excitonic excitation} \label{sec_exc_excitation}
\begin{figure}[t]
\includegraphics[width=\columnwidth]{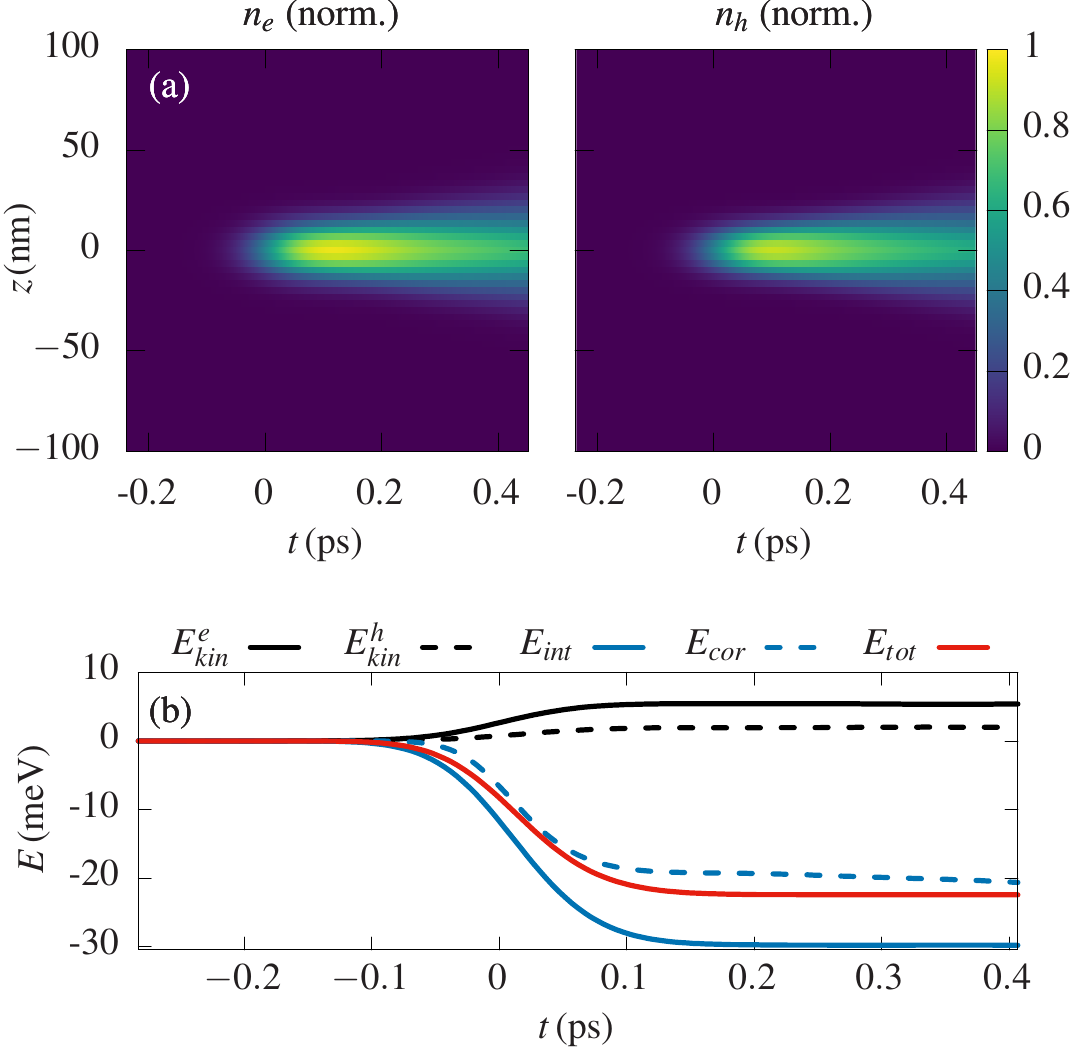}
\caption{Spatiotemporal carrier dynamics in the high-density case after a $100\,$fs pulse with excess energy $E_{\mathrm{ex}}=-24\,$meV. (a) Dynamics of electron and hole distributions and (b) energy contributions.}
\label{fig_5}
\end{figure}
Starting with the excitation at the exciton resonance, we increase the pulse strength, such that on average $N_e\approx 1.7$ electrons are excited. The corresponding density dynamics and energy contributions are shown in Fig.~\ref{fig_5}. We again see that the density is strongly localized at $z=0$ with little movement. Comparing the densities in the high-density [Fig.~\ref{fig_5}(a)] and  low-density case [Fig.~\ref{fig_3}(a)], one observes an increased broadening of the carrier densities with time, which will be further quantified below. The increased broadening of the carrier densities can be understood in terms of Coulomb scattering processes which broaden the momentum distribution. Even though Markovian Coulomb-induced intraband scattering does not exist in a 1D system, on the ultrafast timescales considered here quantum kinetic Coulomb scattering can strongly affect the carrier distributions \cite{Prengel99}. \\
Figure~\ref{fig_5}(b) shows the energy contributions. Comparing the energetic contributions with the energy contributions in the low-density case (Fig.~\ref{fig_4}(a)) one surprisingly sees that most energies do not change considerably. Only the correlation energy is now dominating the interaction energy which shows that the carrier interaction is now dominated by correlations beyond the Hartree-Fock picture. Nevertheless, it is still reasonable to say that the carriers are still in a bound state with $E_{\mathrm{tot}}\approx -22\,$meV. A direct comparison of the spatiotemporal dynamics with a pure Hartree-Fock treatment is presented in App.~\ref{app_theory}.\\
Concerning the spatial broadening of the carrier densities, the interesting question arises whether the transport is still ballistic in view of the time and length scales or already diffusive in view of scattering mechanisms. To answer this question we calculate the variance of the distribution 
	\begin{align*}
	\Delta z^2_i
	=\frac{1}{N_{i}(t)}\int z^2 \cdot n_i(z) dz,
	\end{align*}
where $N_{i}(t)$ ($i=e,h$) is the number of excited particles at time $t$. 
\begin{figure}
\includegraphics[width=\columnwidth]{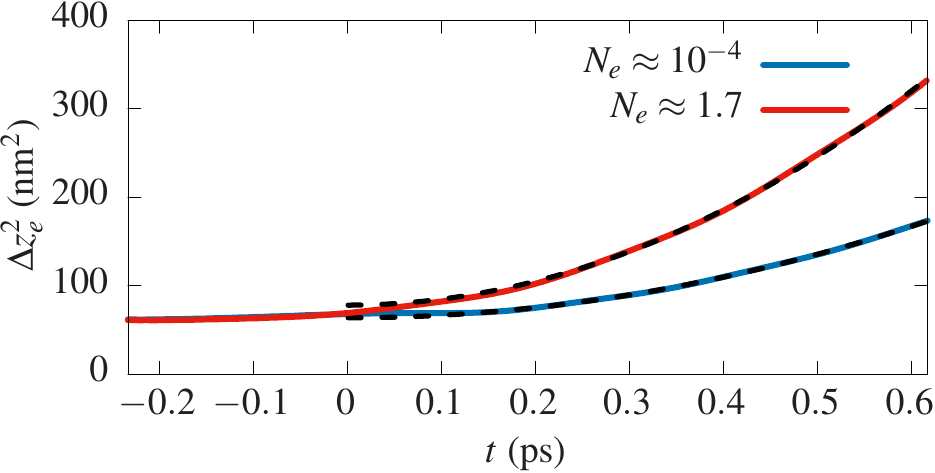}
\caption{Variance of the electronic wave packet as function of time for the low-intensity ($N_e\approx 10^{-4}$, dashed line) and high-intensity ($N_e\approx 1.7$, solid line) case. The black lines indicate quadratic fits $\propto t^2$ to the respective curves.}
\label{fig_6}
\end{figure}
The power dependence of $\Delta z^2_i\propto t^m$ is then an indicator for the transport regime, where $m=1$ corresponds to diffusion and $m=2$ to ballistic motion \cite{Hufnagel01}. The variance of the electronic density for the two pulse intensities is shown in Fig.~\ref{fig_6}. For both excitation regimes one can clearly see the dependence $\Delta z_i^2\propto t^2$, which is confirmed by fits (black dashed lines). Therefore we conclude that the transport is still ballistic (the hole variance shows the same scaling and is therefore not shown here). Nevertheless one can observe a huge increase in ballistic broadening with increased density due to Coulomb correlations; note that a quantum-mechanical speed-up of the ballistic broadening has also been observed as a result of other scattering mechanisms \cite{Rosati14}. This behavior can also be understood by the fact, that for elevated densities the excited carriers are not anymore described by a completely excitonic excitation, but by an admixture of continuum excitations which are spatially less bound (see also Sec.~\ref{sec_excitons}).\\
\subsubsection{Continuum excitation} \label{sec_cont_excitation}
\begin{figure}[t]
\includegraphics[width=\columnwidth]{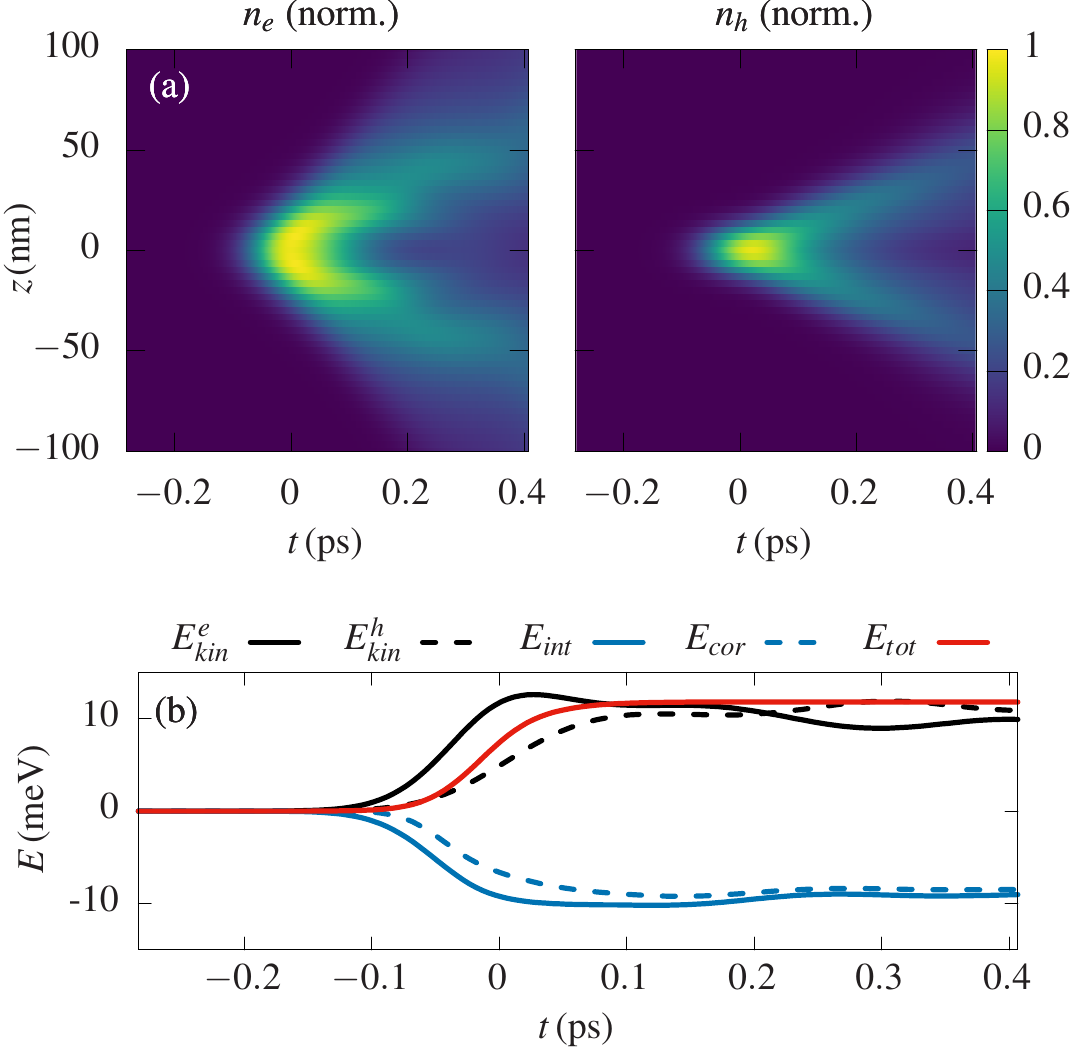}
\caption{Spatiotemporal carrier dynamics after a $100\,$fs pulse with excess energy $E_{\mathrm{ex}}=10\,$meV. (a) Dynamics of electron and hole distributions and (b) energy contributions.}
\label{fig_7}
\end{figure}

\begin{figure}[t]
\includegraphics[width=\columnwidth]{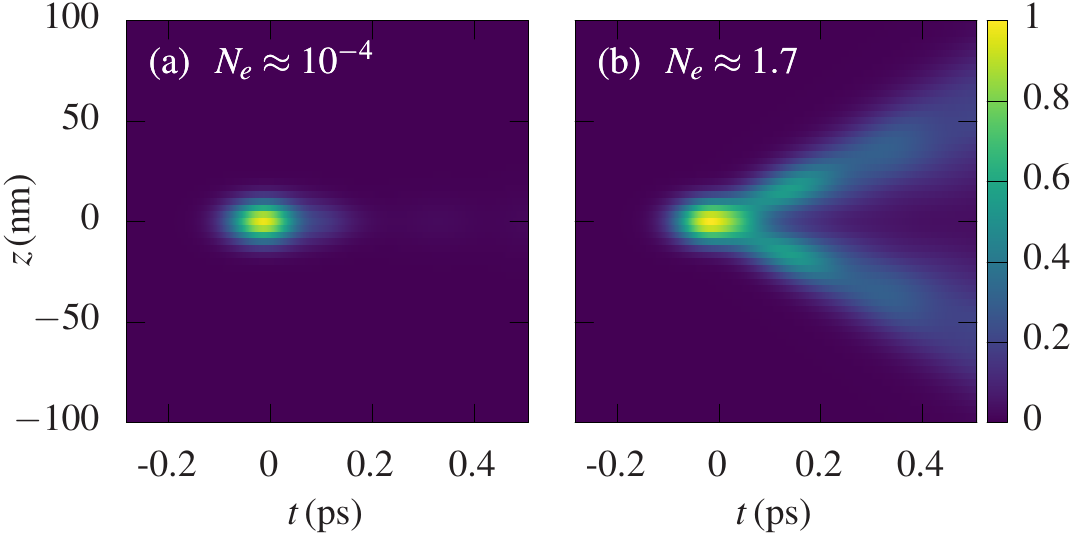}
\caption{Normalized 2-particle density [cf. Eq.(\ref{eq_PL})] for continuum excitation in (a) the low-density limit and (b) the high-density limit.}
\label{fig_8}
\end{figure}

\begin{figure}[t]
\includegraphics[width=\columnwidth]{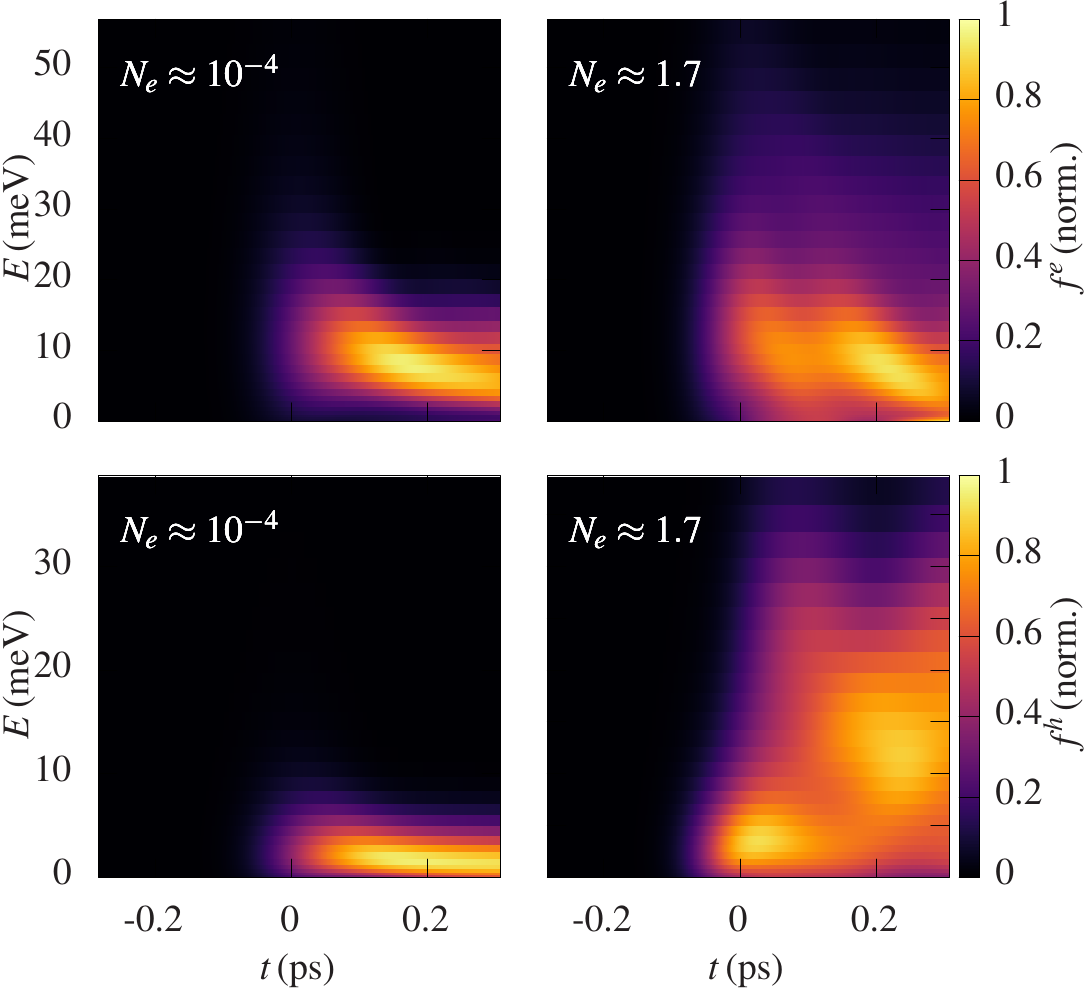}
\caption{Energy distribution of the carrier occupations $f^{e/h}$ as function of time in the low- (left panels) and (b) high-density limit (right panels).}
\label{fig_9}
\end{figure}

We now consider the case of continuum excitation in the high-density limit for an excitation with a pulse strength, such that $N_e\approx 1.7$. The corresponding dynamics of the densities is shown in Fig~\ref{fig_7}(a). For both electrons and holes, we find that wave packets are excited, which travel along the wire. In contrast to the low-intensity limit [Fig.~\ref{fig_3}(b)], now a strong spatial spreading of the densities and an acceleration of the hole wave packet is found.\\
When looking at the corresponding energies, shown in Fig.~\ref{fig_7}(b), the acceleration of holes is directly visible in the kinetic energy $E_{\mathrm{kin}}^h$, which now is on the same level as the kinetic energy of electrons $E_{\mathrm{kin}}^e$. The increase in kinetic energy is compensated by a strong negative interaction energy $E_{\mathrm{int}}$, which for high-densities is dominated by the correlation energy, such that the total energy is only slightly changed in comparison to the low-density case.\\ 
Let us discuss in some more detail, the strong acceleration of holes, which is the most striking difference to the low-density case. The holes are accelerated such that they travel with approximately the same velocity as the electrons. This can already be explained on a Hartree-Fock level where the strong charge density induced by the local excitation leads to a strong electrostatic interaction of the wave packets ultimately forming an ambipolar wave packet effectively reducing the charge density (see App.~\ref{app_theory} for a Hartree-Fock simulation). 
The formation of an ambipolar wave packet as in the present case, where electrons and holes are moving with approximately the same velocity, can be directly mapped by the two-particle density with equal electron and hole position
\begin{align}
	&N_{2P}({z})
	=\sum\limits_{\sigma}\Braket{\hat{\Psi}^{\dagger e}_{\sigma}({z})\hat{\Psi}^{\dagger h}_{\bar{\sigma}}({z})\hat{\Psi}^h_{\bar{\sigma}}({z})
	\hat{\Psi}^e_{\sigma}({z})}
	\label{eq_PL}\\
	&=\frac{1}{V^2}\sum\limits_{\substack{k,k',\sigma\\K,K'}} e^{i(K-K')z}\Braket{\cre{c}_{k'+\frac{K'}{2}\sigma}\cre{d}_{-k'+\frac{K'}{2}\bar{\sigma}}\hat{d}_{-k+\frac{K}{2}\bar{\sigma}}\hat{c}_{k+\frac{K}{2}\sigma}},
	\nonumber
	\end{align}
where $\hat{\Psi}^{\dagger i}_{\sigma}(z)$ are the field operators creating a carrier $i$ at position $z$. Here $k, k'$ describe the relative momenta and $K, K'$ the center-of-mass momenta of the carriers. We plot the two-particle density of Eq.~(\ref{eq_PL}) for the low- and high-density limit in Fig.~\ref{fig_8}, which should be compared to the densities shown in Fig.~\ref{fig_3}(b) and Fig.~\ref{fig_7}(a), respectively.
%The spatial information is inherent in the two-particle coherences because for $K=K'$ the result is spatially homogeneous. 	
In the low-density limit (Fig.~\ref{fig_8}(a)) a fast temporal decay of the two-particle density is visible because of the spatial separation of electron and hole wave packets. In contrast, in the case of high density (Fig.~\ref{fig_8}(b)) one observes a traveling wave packet tracking the motion of carriers directly reflecting the formation of the ambipolar wave packet. \\ 
Another quantification of the acceleration of holes can be gained from the dynamics of the carrier occupations $f^e(\epsilon_k^e)=\Braket{\cre{c}_{{k}}\hat{c}_{{k}}}$, $f^h(\epsilon_k^h)=\Braket{\cre{d}_{{k}}\hat{d}_{{k}}}$ shown in Fig.~\ref{fig_9} for (a) the low-density limit with $N_e\approx 10^{-4}$ and (b) the high-density limit with $N_e\approx 1.7$. In the low-density limit we observe essentially an excitation at $k$-values corresponding to the excess energy $k_0=\sqrt{2\mu E_{\mathrm{ex}}}/\hbar$ ($\epsilon_{k_0}^e\approx E_{\mathrm{ex}}> \epsilon_{k_0}^h$) distributed between electrons and holes. During the pulse, due to the renormalization by Coulomb effects, the energy distribution is smeared out. \\
In contrast, when looking at the dynamics in the high-density case, the electronic distribution (upper panel) is mostly smeared out by correlation effects and the hole distribution  shows a strong acceleration from $E\approx 3\,$meV to $E\approx 12\,$meV, resulting in the acceleration of the hole wave packets observed in Fig.~\ref{fig_7}(a). Additionally a strong broadening of the distribution is observed which is even stronger than in the case of electrons because the Coulomb scattering of holes is facilitated by their flatter band structure. The strong energetic broadening of the distributions leads to the enhanced spatial spreading, because a broader range of wave-packet velocities contributes to the wave packet.

\section{Excitonic picture}\label{sec_excitons}
To discriminate more clearly between the excitonic occupations and the continuum contributions for the different excitations, we now analyze our results using an excitonic picture. In this picture all quantities are transformed into the two-particle picture defined by the excitonic eigenfunctions \cite{Siantidis01,Katsch18}. This will be particulary helpful for the high-density limit, where the distinction between exciton and continuum carriers becomes questionable.
The exciton occupations can be defined as 
\begin{align*}
\Braket{\cre{Y}_x\hat{Y}_x}
&:=\sum\limits_{{q,q}',\sigma}g_{{q},K-q}^x
\left(g_{{q}',K'-q'}^x\right)^*\\
&\cdot\Braket{\cre{c}_{{q},\sigma}\cre{d}_{{K-q},\bar{\sigma}}\hat{d}_{{K-q'},\bar{\sigma}}\hat{c}_{{q}',\sigma}},
\end{align*}
with $x=(n,{K})$ consisting of the center-of-mass momentum ${K}$ and the hydrogen-like quantum number $n$ (1s,..., continuum). $\hat{Y}_x (\cre{Y}_x)$ is the exciton annihilation (creation) operator. The expansion coefficients $g_{{q,K-q}}^x$ describe the transformation from the free-particle states to the exciton states with the relative momentum $q$ (see App.~\ref{app_exc}). Having done that, we compute the fraction of 1s-excitons and continuum excitons using
	\begin{align*}
	N^{x}_{tot}=\sum\limits_x \Braket{\cre{Y}_x\hat{Y}_x}=N_{1s}+N_C
	\end{align*}
with $N_{1s}=\sum\limits_{{K}} \Braket{\cre{Y}_{1s,{K}}\hat{Y}_{1s,{K}}}$. Thereby one can distinguish between excited bound electron-hole pairs and electron-hole pairs within the excitonic continuum which essentially behave as independent particles. We here consider all states above the $1s$ states as a continuum, because higher excitonic states merge with the continuum excitations (cf. Fig.~\ref{fig_2}). 

\begin{figure}
 \includegraphics[width=\columnwidth]{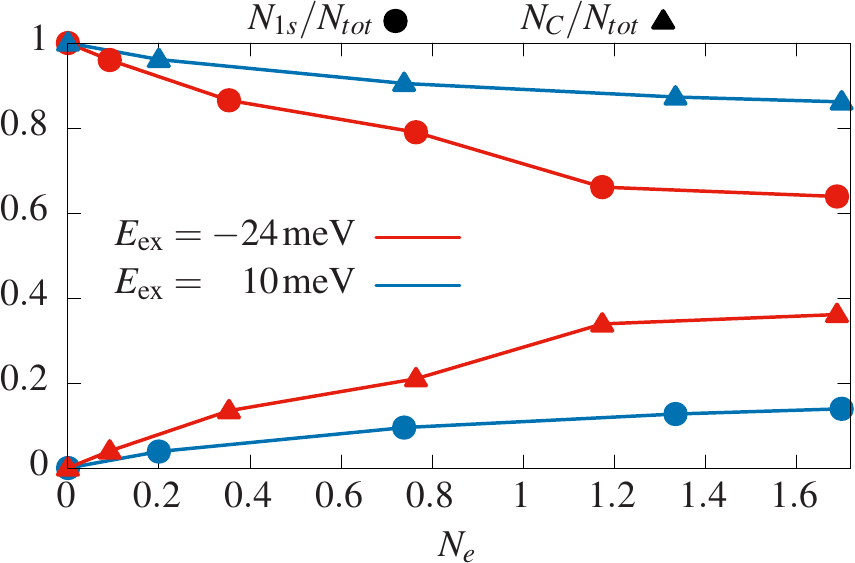}
 \caption{Final excitonic (circles) and continuum (triangles) occupation probabilities for different carrier densities for $E_{\mathrm{ex}}=-24\,$meV (red symbols) and $E_{\mathrm{ex}}=10\,$meV (blue symbols).}
 \label{fig_10}
\end{figure}

The stationary values of $N_{1s}$ and $N_C$ (normalized to the total number of two-pair states) at the end of the simulation ($t\approx 0.45\,$ps) are shown in Fig.~\ref{fig_10} as a function of the excitation power quantified by the number of electrons $N_e$. One can directly see, that the behavior for the two excess energies is opposed to each other. In the low-density case there are only excitonic carriers for $E_{\mathrm{ex}}=-24\,$meV (red symbols) and only continuum carriers for $E_{\mathrm{ex}}=10\,$meV (blue symbols). For increasing densities the excitonic and continuum carriers start to mix. For the excitonic excitation, this is expected, because the excitonic resonance becomes screened for increased carrier densities. In the case of continuum excitation the Coulomb interaction between the finite densities leads to the fact, that the wave packets, which are initially of free-carrier character, interact with each other and thereby acquire excitonic nature. It is worth noting, that the mixing of continuum and excitonic carriers is more strongly influenced by the density for the excitonic excitation than for the continuum excitation, such that at $N_e\approx 1.7$ the carriers are only $60\%$ excitonic.

\begin{figure}
\includegraphics[width=\columnwidth]{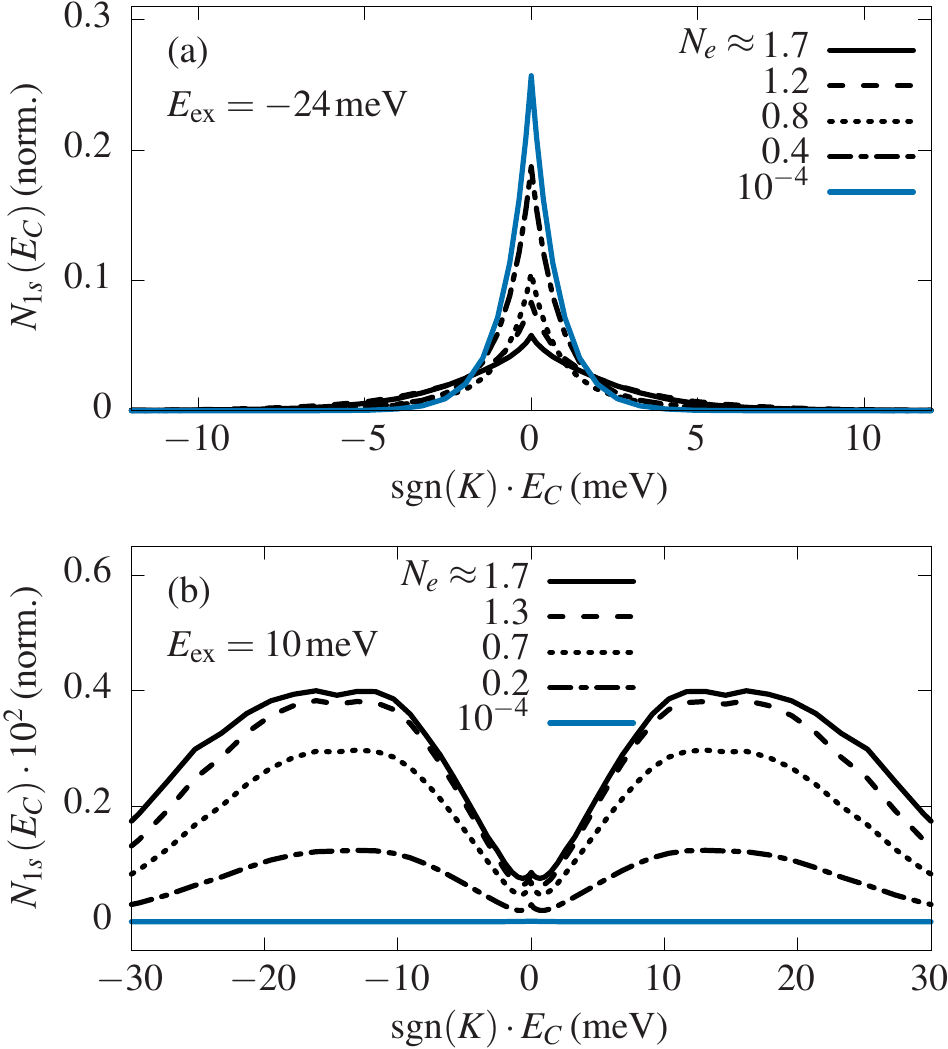}
\caption{Normalized 1s exciton distribution as function of center-of-mass energy at the end of the simulation for (a) excitonic and (b) continuum excitation. The black solid line corresponds to the high-density limit and the blue solid line to the low-density limit.}
\label{fig_11}
\end{figure}

We further look at the exciton occupations within the 1s band
	\begin{align*}
	N_{1s}\left(E_{C}\right)=\Braket{\cre{Y}_{1s,{K}(E_C)}\hat{Y}_{1s,{K}(E_C)}},
	\end{align*}
where $E_{C}$ denotes the center-of-mass energy. The final distribution in the two excitation scenarios is shown in Fig.~\ref{fig_11} for different excitation strengths denoted by the number of particles $N_e$. In Fig.~\ref{fig_11}(a) we consider the excitonic excitation. In the low-density limit the width of the distribution (blue line) is exclusively determined by the spatial localization of the exciting pulse. For increased intensity (increasing $N_e$) the distribution successively broadens, such that a considerable fraction of excitons with finite center-of-mass momentum is present. This energetic broadening to higher center-of-mass energies is in agreement with the enhanced ballistic spreading for higher carrier densities together with the admixture of continuum carriers shown in Fig.~\ref{fig_10}. For the continuum excitation presented in Fig.~\ref{fig_11}(b), we find a much broader distribution to start with, which is centered around a finite center-of-mass energy representing the fact, that moving ambipolar wave packets are traveling through the quantum wire. With increasing excitation strength the finite momentum successively builds up from the low-density limit, where the center-of-mass momentum is essentially zero and the 1s occupation is vanishing. Thereby, it is the acceleration of holes, which leads to the fact that the absolute values of electron and hole momenta are not equal anymore and excitonic occupations with finite center-of-mass momentum $K=k_e+k_h$ build up.
\section{Conclusion}	 
\label{sec_concl}
In summary, we have discussed the impact of Coulomb effects on the ultrafast spatio-temporal carrier dynamics in semiconductors. As an example, we have considered the dynamics in a semiconductor quantum wire. For this, we have performed calculation in a configuration interaction like approach. By restricting ourselves to at most two electron-hole pairs, we obtain an exact model ready to scrutinize the Coulomb effects and correlations beyond Hartree-Fock level. This allowed us to examine the influence of the Coulomb interaction on the exciton dynamics in particular for different carrier densities. While in the low-density limit, where only few carriers are excited, the exciton and continuum states could be well separated, for higher excitation density Coulomb correlations lead to mixing of those. \\
These effects are also seen in the spatio-temporal dynamics of the excited carriers, which behave qualitatively different in the low- and high-density limit. The excitation resonant to the exciton leads to mostly stationary carriers, which for higher density still move ballistically, but much faster away from excitation region resulting in an ultrafast spatial spreading. For excitations in the continuum, two mostly independent electron and hole wave packets were formed traveling along the quantum wire. In contrast, for higher densities a strong acceleration of the hole wave packet led to a formation of an ambipolar wave packet. This wave packet can be dominantly characterized by carriers within the excitonic continuum and a fraction of excitons with finite center-of-mass momentum.\\
Our studies give important insights in the dynamics of excitons in low-dimensional semiconductors and will help in the development of exciton-based devices in semiconductor technology. 
\section*{Acknowledgements}
 F.~L. and D.~E.~R. acknowledge financial support by the Deutsche Forschungsgemeinschaft (DFG) through the project %\textit{Simulation der optisch induzierten und r\"aumlich aufgel\"osten Ladungstr\"agerdynamik in zweidimensionalen Halbleitern} 
406251889 (RE 4183/2-1). We also thank V.~M.~Axt for fruitful discussions. 
	\appendix

\section{Comparison to other theoretical treatments} \label{app_theory}
Here, we briefly compare our wave-function based CI approach to density matrix approaches, which are commonly used to treat photoexcited semiconductors \cite{Rossi02,Kira06} and - as a specific example - to a Hartree-Fock calculation.\\
We remark that our CI approach is able to describe quantities up to a biexciton-occupation exactly, which is described by an eight-operator expectation value. In perturbative approaches those expectation values are usually expanded in products of lower expectation values and correlations of a certain order are neglected. While such a decoupling scheme can, e.g., be unambiguously truncated in finite orders of the electric field for nonlinear spectroscopy \cite{Axt94}, it is hard to define a truncation scheme for high excitation densities, especially when only a few carriers are excited, i.e., for strong spatial localizations, where correlation effects should dominate. Mean-field treatments, like the Hartree-Fock factorization, which are usually the basis of density-matrix approaches, are not justified in those cases. An additional advantage of the CI method in comparison to the aforementioned density matrix approaches is the fact that the evaluation of the Schr\"o{dinger} equation will result in linear differential equations, which are numerically much more stable than the nonlinear equations of motion of a density-matrix based decoupling scheme. The CI approach is therefore powerful in the context of strongly localized excitation since here only a few particles are present in the system and their correlation is of crucial importance. Nevertheless it is mandatory, that the number of particles is known and fixed throughout the simulation. This makes the treatment of effects like Auger-recombination or impact-ionization and the treatment of bosonic particles difficult.

\begin{figure}[t]
\includegraphics[width=\columnwidth]{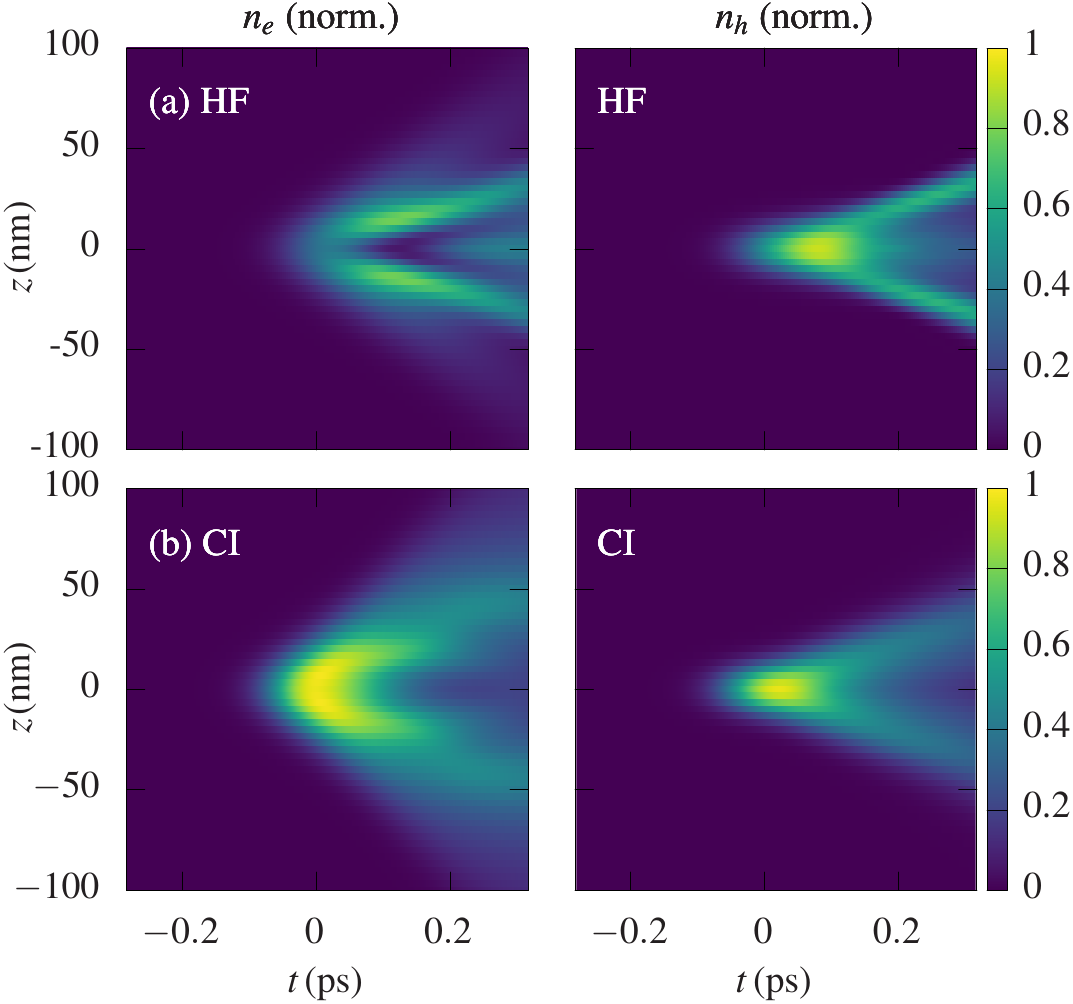}
\caption{Electronic (left column) and hole (right column) densities for continuum excitation in the high-density limit using (a) a Hartree-Fock (HF) simulation and (b) the wave-function based CI simulation (cf. Fig~\ref{fig_7}(a)).}
\label{fig_Continuum_Hartree-Fock_compare}
\end{figure}

%\subsection*{Comparison to a Hartree-Fock Simulation}\label{app_Hartree-Fock}
It is interesting to directly compare our approach to a Hartree-Fock (HF) calculation. The equations of motion in this case read
\begin{align*}
\frac{d}{dt}f_{k,k'}^e=
&\frac{i}{\hbar}\sum\limits_{k''}\left[\mathcal{E}_{k,k''}^ef_{k'',k'}^e-f_{k,k''}^e\mathcal{E}_{k'',k'}^e\right]\\
&-\frac{i}{\hbar}\sum\limits_{k''}\left[\mathcal{U}_{k'',k}^*p_{k'',k'}-p_{k'',k}^*\mathcal{U}_{k'',k'}\right]
\end{align*}
\begin{align*}
\frac{d}{dt}f_{k,k'}^h=
&\frac{i}{\hbar}\sum\limits_{k''}\left[\mathcal{E}_{k,k''}^hf_{k'',k'}^h-f_{k,k''}^h\mathcal{E}_{k'',k'}^h\right]\\
&-\frac{i}{\hbar}\sum\limits_{k''}\left[\mathcal{U}_{-k,-k''}^*p_{-k',-k''}-p_{-k,-k''}^*\mathcal{U}_{-k',-k''}\right]
\end{align*}
\begin{align*}
\frac{d}{dt}p_{k,k'}=
&-\frac{i}{\hbar}\sum\limits_{k''}\left[\mathcal{E}_{-k'',-k}^hp_{k'',k'}+p_{k,k''}\mathcal{E}_{k'',k'}^e\right]\\
&+\frac{i}{\hbar}\sum\limits_{k''}\left[\left(\delta_{k'',k}-f_{-k'',-k}^h\right)\mathcal{U}_{k'',k'}-\mathcal{U}_{k,k''}f_{k'',k'}^e\right]
\end{align*}
with $f_{k,k'}^e=f_{k,k',\sigma}^e=\Braket{\cre{c}_{k\sigma}\hat{c}_{k'\sigma}}$, $f_{k,k'}^h=f_{k,k',\sigma}^h=\Braket{\cre{d}_{k\sigma}\hat{d}_{k'\sigma}}$, $p_{k,k'}=p_{k,k',\sigma}=\Braket{\hat{d}_{-k\bar{\sigma}},\hat{c}_{k'\sigma}}$, the renormalized energies
	\begin{align*}
	\mathcal{E}_{k,k'}^{e/h}
	=&\epsilon_k^{e/h}\delta_{k,k'}-\sum\limits_q V_q f_{k+q,k'+q}^{e/h}\\
	&+2V_{k-k'}\sum\limits_q\left(f_{k+q,k'+q}^{e/h}-f_{k+q,k'+q}^{h/e}\right)
	\end{align*}
and the renormalized fields
$$\mathcal{U}_{k,k'}=E_{k,k'}(t)+\sum\limits_q V_q p_{k+q,k'+q}$$ \cite{Rossi02}.
For this, we show the continuum excitation in the high-density limit for CI and Hartree-Fock in Fig.~\ref{fig_Continuum_Hartree-Fock_compare}. The upper panel shows the Hartree-Fock calculations and the lower one shows the CI calculations (these were already shown in Fig.~\ref{fig_7} in Sec.~\ref{sec_cont_excitation}). For both calculations one finds that electron and hole wave packets are formed, which travel along the wire with the same speed. In other words, the general trend of the formation of an ambipolar wave packet is already described on the Hartree-Fock level underlining the interpretation, that the driving force for this formation is the classical electrostatic attraction between electron and hole. Nevertheless the spatial broadening of the densities is underestimated on the Hartree-Fock level, because quantum kinetic scattering is not incorporated. For the same reason there is a density at $z=0$ in Hartree-Fock building up after the pulse has ended, because the renormalized electric field does not dephase in the Hartree-Fock treatment. We conclude that the Hartree-Fock approximation gives a qualitatively correct prediction of the ambipolar wavepacket and therefore might be a valid choice for dynamical calculations of wave packets.

\begin{figure}[]
\includegraphics[width=\columnwidth]{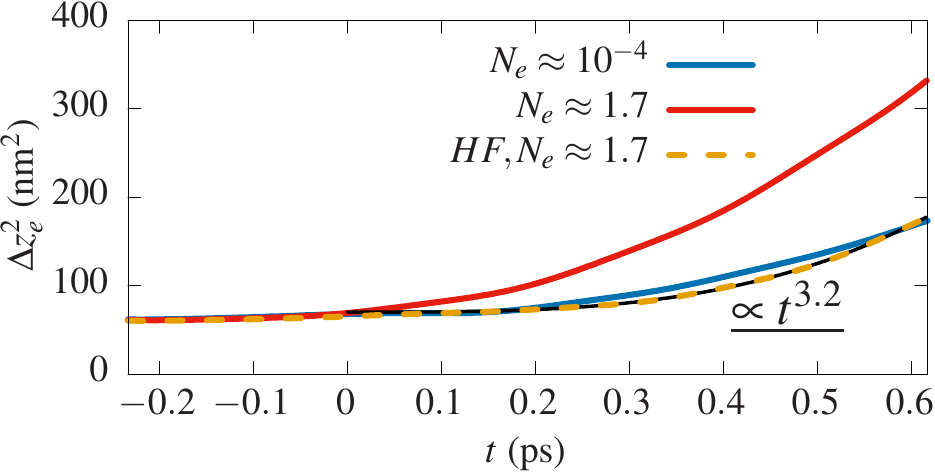}
\caption{Wave packet variance $\Delta z_e^2$ as function of time (cf. Fig.~\ref{fig_6}) now including additionally Hartree-Fock simulation (orange dashed line). A polynomial fit is shown as black solid line.}
\label{fig_Excitonic_Hartree-Fock_compare}
\end{figure}

Finally, we want to study, if also the ballistic nature of the transport in case of the excitonic excitation is already captured in a Hartree-Fock calculation. Therefore, we show in Fig.~\ref{fig_Excitonic_Hartree-Fock_compare} the dynamics of $\Delta z_e^2$ for an excitonic excitation and compare the CI simulation (see also Fig.~\ref{fig_6}) with the results from the Hartree-Fock calculation given in orange (dashed). The Hartree-Fock case strongly underestimates the spatial spreading of the wave packet and also falsely predicts a super-ballistic behavior with $\Delta z_e^2\propto t^{3.2}$. Therefore in this case the Hartree-Fock approximation fails to correctly describe the dynamical behavior of the exciton.

In summary, the comparison with a Hartree-Fock treatment underlines the fact, that a correlation-expansion scheme is a valid method for systems where energetic continua play a role as in the case of continuum excitation. In the case of excitonic excitation there is the center-of-mass continuum, but nevertheless no continuum in relative coordinates.

\section{Excitonic eigenspace} \label{app_exc}
We here briefly summarize the transformation to the exciton eigenspace. We consider the ansatz for the excitonic eigenfunctions $\ket{\varphi^x}=\sum\limits_{k_1,k_2}g_{k_1,k_2}^x\cre{c}_{k_1\sigma_1}\cre{d}_{k_2 {\sigma_2}}
	\ket{0}$. 
	This results in the Wannier equation 
		\begin{align*}	\sum\limits_{{q}}\left[(\epsilon_{{k}_1}^e+\epsilon_{k_2}^h)\delta_{{q},0}
	-V({q})\right]g^{x}_{{k}_1+{q},{k}_2-{q}}=E_{x}g_{{k}_1,{k}_2}^{x}
	\end{align*}
after projecting the system Hamiltonian on $\ket{\varphi^x}$.
The solution reads with $x=n,K$
	\begin{align*}g^{n,K}_{k_1,k_2}
	&=\delta_{k_1+k_2,K}\tilde{\phi}_n\left(\frac{m_h}{M}k_1-\frac{m_e}{M}k_2\right)\\
	&=\delta_{k_2,K-k_1}\tilde{\phi}_n\left(k_1-\frac{m_e}{M}K\right),
	\end{align*}
$K=k_1+k_2$ being the center-of-mass momentum of the exciton, $M=m_e+m_h$ and
$\tilde{\phi}_n$ determined by the solution of the effective hydrogen problem
	\begin{align*}
	\sum\limits_{q}\left[\frac{\hbar^2k^2}{2\mu}\delta_{q,0}-V(q)\right]\tilde{\phi}_n(k+q)=\epsilon_n\tilde{\phi}_n(k)
	\end{align*}
with the reduced mass $\mu=\frac{m_em_h}{m_e+m_h}$.\\
Within this basis one can obtain, e.g., the excitonic polarization as
	\begin{align*}
	\cre{Y}_{n,K}
	&=\sum\limits_{k_1,k_2} g_{k_1,k_2}^{n,K}\cre{c}_{k_1\sigma}\cre{d}_{k_2\bar{\sigma}}\\
	&=\sum\limits_{k_1}\tilde{\phi}_n\left(k_1-\frac{m_e}{M}K\right)\cre{c}_{k_1\sigma}\cre{d}_{K-k_1\bar{\sigma}},
	\end{align*}
where we defined the optically active excitonic state.

\end{document}